\documentclass[12pt]{article}

\usepackage{graphicx}
\usepackage{amsmath}
\usepackage{amssymb}
\usepackage{amsthm}
\usepackage{color}
\usepackage{lscape}
\usepackage{soul}
\usepackage{multirow}

\renewcommand{\thefootnote}{\fnsymbol{footnote}}

\begin{document}

\begin{flushright}
CYCU-HEP-14-05  \\
EPHOU-14-013
\end{flushright}

\vspace{4ex}

\begin{center}

{\LARGE\bf Gauge Origin of Discrete Flavor Symmetries in Heterotic Orbifolds}

\vskip 1.4cm

{\large  
Florian Beye$^{1}$\footnote{Electronic address: fbeye@eken.phys.nagoya-u.ac.jp},
Tatsuo Kobayashi$^{2}$\footnote{Electronic address: kobayashi@particle.sci.hokudai.ac.jp}
and
Shogo Kuwakino$^{3}$\footnote{Electronic address: kuwakino@cycu.edu.tw}
}
\\
\vskip 1.0cm
{\it $^1$Department of Physics, Nagoya University, Furo-cho, Chikusa-ku, Nagoya 464-8602, Japan} \\
{\it $^2$Department of Physics, Hokkaido University, Sapporo 060-0810, Japan} \\
{\it $^3$Department of Physics, Chung-Yuan Christian University, 200, Chung-Pei Rd. Chung-Li,320, Taiwan} \\

\vskip 3pt
\vskip 1.5cm

\begin{abstract}
We show that non-Abelian discrete symmetries in orbifold string models have a gauge origin. This can be understood when looking at the vicinity of a symmetry enhanced point in moduli space. At such an enhanced point, orbifold fixed points are characterized by an enhanced gauge symmetry. This gauge symmetry can be broken to a discrete subgroup by a nontrivial vacuum expectation value of the K\"ahler modulus $T$. Using this mechanism it is shown that the $\Delta(54)$ non-Abelian discrete symmetry group originates from a $SU(3)$ gauge symmetry, whereas the $D_4$ symmetry group is obtained from a $SU(2)$ gauge symmetry. 
\end{abstract}

\end{center}

\newpage

\setcounter{footnote}{0}
\renewcommand{\thefootnote}{\arabic{footnote}}

\newpage

\section{Introduction}

It is important to understand the flavor structure of the standard model of particle physics. Quark and lepton masses are hierarchical. Two of the mixing angles in the lepton sector are large, while the mixing angles in the quark sector are suppressed, except for the Cabibbo angle. Non-Abelian discrete flavor symmetries may be useful to understand this flavor structure. Indeed, many works have considered field-theoretical model building with various non-Abelian discrete flavor symmetries (see \cite{Altarelli:2010gt,Ishimori:2010au,King:2013eh} for reviews).

Understanding the origin of non-Abelian flavor symmetries is an important issue we have to address. It is known that several phenomenologically interesting non-Abelian discrete symmetries can be derived from string models\footnote{In \cite{Continuous}, field theoretical models where non-Abelian discrete groups are embedded into non-Abelian gauge groups are considered.}. In intersecting and magnetized D-brane models, the non-Abelian discrete symmetries $D_4$, $\Delta(27)$ and $\Delta(54)$ are realized \cite{Abe:2009vi,Intersecting,Hamada:2014hpa,Abe:2014nla}. Also, their gauge origins have been studied \cite{Intersecting}. In heterotic orbifold compactifications \cite{orbifold, heteroph, Katsuki:1989bf, Kobayashi:2004ud, Buchmuller, Kim, Lebedev:2006kn, Blaszczyk:2009in, Nibbelink:2013lua} (also see a review \cite{Nilles:2008gq}), non-Abelian discrete symmetries appear due to geometrical properties of orbifold fixed points and certain properties of closed string interactions \cite{Kobayashi:2006wq}. First, there are permutation symmetries of orbifold fixed points. Then, there are string selection rules which determine interactions between  orbifold sectors. The combination of these two kinds of discrete symmetries leads to a non-Abelian discrete symmetry. In particular, it is known that the $D_4$ group emerges from the one-dimensional orbifold $S^1/Z_2$, and that the $\Delta(54)$ group is obtained from the two-dimensional orbifold $T^2/Z_3$. The phenomenological applications of the string-derived non-Abelian discrete symmetries are analyzed e.g. in \cite{Ko:2007dz}.

In this paper we point out that these non-Abelian discrete flavor symmetries originate from a gauge symmetry. To see this, we consider a heterotic orbifold model compactified on some six-dimensional orbifold. The gauge symmetry $G_{{\rm gauge}}$ of this orbifold model is, if we do not turn on any Wilson lines, a subgroup of $E_8 \times E_8$ which survives the orbifold projection. In addition, from the argument in \cite{Kobayashi:2006wq}, we can derive a non-Abelian discrete symmetry $G_{\rm discrete}$. Then, the effective action of this model can be derived from $G_{{\rm gauge}} \times G_{\rm discrete}$ symmetry invariance\footnote{Here we do not consider the $R$-charge invariance since this is not relevant to our discussion.}. However, this situation slightly changes if we set the model to be at a symmetry enhanced point in moduli space. At that special point, the gauge symmetry of the model is enlarged to ${G_{{\rm gauge}} \times G_{\rm enhanced}}$, where $G_{\rm enhanced}$ is a gauge symmetry group. The maximal rank of the enhanced gauge symmetry $G_{\rm enhanced}$ is six, because we compactify six internal dimensions. At this specific point in moduli space, orbifold fixed points are characterized by gauge charges of $G_{\rm enhanced}$, and the spectrum is extended by additional massless fields charged under $G_{\rm enhanced}$. Furthermore, the K\"ahler moduli fields $T$ in the untwisted sector obtain $G_{\rm enhanced}$-charges and a non-zero vacuum expectation value (VEV) of $T$ corresponds to a movement away from the enhanced point. This argument suggests the possibility that the non-Abelian discrete symmetry $G_{{\rm discrete}}$ is enlarged to a continuous gauge symmetry $G_{{\rm enhanced}}$ at the symmetry enhanced point. In other words, it suggests a gauge origin of the non-Abelian discrete symmetry. Moreover, the group $G_{{\rm enhanced}}$ originates from a larger non-Abelian gauge symmetry that exists before the orbifolding. We will show this explicitly in the following.

\section{Gauge origin of non-Abelian discrete symmetry}

In this section we demonstrate the gauge origin of non-Abelian discrete symmetries in heterotic orbifold models. We concentrate on the phenomenologically interesting non-Abelian discrete symmetries $D_4$ and $\Delta(54)$ which are known to arise from orbifold models. 

\subsection{$D_4$ non-Abelian discrete symmetry}

First, we study a possible gauge origin of the $D_4$ non-Abelian discrete symmetry. This symmetry is associated with the one-dimensional $S^1/Z_2$ orbifold. We consider the heterotic string on a factorizable six-dimensional orbifold which includes $S^1/Z_2$. The coordinate corresponding to this one dimension is denoted by $X$. It suffices to discuss only the left-movers in order to develop our argument. Let us start with the discussion on $S^1$ without the $Z_2$ orbifold. There is always a $U(1)$ symmetry associated with the current $H = i\partial X$. At a specific point in the moduli space, i.e. at a certain radius of $S^1$, two other massless vector bosons appear and the gauge symmetry is enhanced from $U(1)$ to $SU(2)$. Their currents are written as 
\begin{eqnarray}
E_{\pm } = e^{\pm i\alpha X},
\end{eqnarray}
where $\alpha=\sqrt 2$ is a simple root of the $SU(2)$ group. These currents, $H$ and $E_{\pm }$, satisfy the $su(2)$ Kac-Moody algebra.

Now, let us study the $Z_2$ orbifolding $X \rightarrow -X$. The current $H = i \partial X$ is not invariant under this reflection and the corresponding $U(1)$ symmetry is broken. However, the linear combination $E_{+} + E_{-}$ is $Z_2$-invariant and the corresponding $U(1)$ symmetry remains on $S^1/Z_2$. Thus, the $SU(2)$ group breaks down to $U(1)$ by orbifolding. Note that the rank is not reduced by this kind of orbifolding.
It is convenient to use the following basis, 
\begin{eqnarray}
H' &=& i\partial X' = \frac{1}{\sqrt 2} \left( E_{+} + E_{-} \right), \\
E'_{\pm } &=& e^{\pm i\alpha X'} = \frac{1}{ \sqrt 2} H \mp \frac12 \left( 
 E_{+} - E_{-}  \right).
\end{eqnarray}
The introduction of the boson field $X'$ is justified because $H'$ and $E'_{\pm}$ satisfy the same operator product expansions (OPEs) as the original currents $H$ and $E_\pm$. The invariant current $H'$ corresponds to the $U(1)$ gauge boson. The $E'_{\pm}$ transform as
\begin{align} \label{Z2P}
E'_{\pm} \rightarrow - E'_{\pm }
\end{align}
under the $Z_2$ reflection and correspond to untwisted matter fields $U_1$ and $U_2$ with $U(1)$ charges $\pm \alpha$. 
In addition, there are other untwisted matter fields $U$ which have vanishing $U(1)$ charge, but are charged under an unbroken subgroup of $E_8 \times E_8$.

From \eqref{Z2P}, it turns out that the $Z_2$ reflection is represented by a shift action in the $X'$ coordinate, 
\begin{align}
X' \rightarrow X' + 2 \pi \frac{ w}{2},
\end{align}
where $w=1/\sqrt 2$ is the fundamental weight of $SU(2)$. That is, the $Z_2$-twisted orbifold on $X$ is equivalent to a $Z_2$-shifted orbifold on $X'$ with the shift vector $s = w/2$ (see e.g., \cite{Dijkgraaf:1987vp}). In the twist representation, there are two fixed points on the $Z_2$ orbifold, to each of which corresponds a twisted state. Note that the one-dimensional bosonic string $X$ with the $Z_2$-twisted boundary condition has a contribution of $h=1/16$ to the conformal dimension. In the shift representation, the two twisted states can be understood as follows. Before the shifting, $X'$ also represents a coordinate on $S_1$ at the enhanced point, so the left-mover momenta $p_L$ lie on the momentum lattice 
\begin{align}
\Gamma_{SU(2)} \; \cup \;  (\Gamma_{SU(2)} + w), 
\end{align}
where $\Gamma_{SU(2)}$ is the SU(2) root lattice, $\Gamma_{SU(2)} \equiv n \alpha$ with integer $n$. Then, the left-mover momenta in the $Z_2$-shifted sector lie on the original momentum lattice shifted by the shift vector $s = w/2$, i.e. 
\begin{eqnarray}
(\Gamma_{SU(2)} + \frac{w}{2}) \; \cup \;  (\Gamma_{SU(2)} + \frac{3w}{2}).
\end{eqnarray}
Thus, the shifted vacuum is degenerate and the ground states have momenta $p_L = \pm \alpha/4$. These states correspond to charged matter fields $M_1$ and $M_2$. Note that $p^2_L/2 = 1/16$, which is exactly the same as the conformal dimension $h=1/16$ of the twisted vacuum in the twist representation. Indeed, the twisted states can be related to the shifted states by a change of basis \cite{Dijkgraaf:1987vp}. Notice that the twisted states have no definite $U(1)$ charge, but the shifted states do. Table \ref{Tab:FieldZ2} shows corresponding matter fields and their $U(1)$ charges.

\begin{table}[t]
\begin{center}
\begin{tabular}{|c|c|c|c|c|}
\hline
Sector & Field & $U(1)$ charge  & $Z_4 $ charge  \\
\hline
\hline
U & $U$ & $ 0  $ & $0$   \\  
\hline
U & $U_1$ & $ \alpha  $  & $0$ \\  
\hline
U & $U_2$ & $ - \alpha  $ & $0$   \\  
\hline
T & $M_1$ & $ \frac{\alpha}{4} $ & $ \frac{1}{4}$ \\  
\hline
T & $M_2$ & $ - \frac{\alpha}{4} $ & $- \frac{1}{4}$    \\  
\hline
\end{tabular}
\caption[smallcaption]{Field contents of $U(1) \rtimes Z_2$ model from $Z_2$ orbifold. $U(1)$ charges are shown. Charges under the $Z_4$ unbroken subgroup of the $U(1)$ group are also shown.}
\label{Tab:FieldZ2}
\end{center}
\end{table}

From Table \ref{Tab:FieldZ2}, we find that there is an additional $Z_2$ symmetry of the (would-be-massless) matter contents: Transforming the $U(1)$-charges $q$ as
\begin{align}
q \to - q,
\end{align}
while at the same time permuting the fields as $U_1 \leftrightarrow U_2$ and $M_1 \leftrightarrow M_2$ maps the spectrum onto itself. The action on the $U_i$ and $M_i$ fields is described by the $2\times 2$ matrix  
 \begin{align} \label{Z2}
&\left( 
\begin{array}{cc}
 0 & 1 \\
 1 & 0  \\
\end{array} 
\right).
\end{align}
This $Z_2$ symmetry does not commute with the $U(1)$ gauge symmetry and it turns out that one obtains a symmetry of semi-direct product structure, $U(1) \rtimes Z_2$.

In the twist representation, this model contains the K\"ahler moduli field $T$, which corresponds to the current $H$ and is charged under the $U(1)$ group. In the shift representation, the field $T$ is described by the fields $U_i$ as
\begin{align}
T = \frac{1}{\sqrt{2}} ( U_1 + U_2  ) \ .
\end{align}
Now we consider the situation where our orbifold moves away from the enhanced point by taking a specific VEV of the K\"ahler moduli field $T$ which corresponds to the VEV direction 
\begin{align}
\langle U_1 \rangle = \langle U_2 \rangle. 
\end{align}
Note that this VEV relation maintains the $Z_2$ discrete symmetry \eqref{Z2}. Moreover, since the fields $U_1$ and $U_2$ are charged under the $U(1)$ gauge symmetry and due to the presence of the $M_i$ fields, the VEV breaks $U(1)$ down to a discrete subgroup $Z_4$. The $Z_4$ charge is $1/4$ for $M_1$ and $-1/4$ for $M_2$ as listed in Table \ref{Tab:FieldZ2}. Written as a $2 \times 2$ matrix, the $Z_4$ action is described by
\begin{align} \label{Z4}
&\left( 
\begin{array}{cc}
 i & 0 \\
 0 & -i \\
\end{array} 
\right).
\end{align}
The matrices \eqref{Z2} and \eqref{Z4} are nothing but the generators of $D_4 \simeq  Z_4  \rtimes Z_2 $. After the VEV, the field $U$ transforms as the trivial singlet ${\bf 1}$, and $(M_1, M_2)$ forms a ${\bf 2}$ representation under the $D_4$ group. This reproduces the known result for a general radius of $S_1$ \cite{Kobayashi:2006wq}. The pattern of symmetry breaking we have shown here is summarized as follows:
\begin{align} 
SU(2) \  \xrightarrow[{\rm orbifolding}]{} U(1) \rtimes Z_2 \  \xrightarrow[\langle T \rangle ]{}  D_4 \ . 
\end{align}
The other VEV directions of $U_1$ and $U_2$ break $U(1) \rtimes Z_2$ to $Z_4$.

\subsection{$\Delta(54)$ non-Abelian discrete symmetry} \label{delta}

Next, we consider the two-dimensional $T^2/Z_3$ orbifold case which is associated with the $\Delta(54)$ non-Abelian discrete symmetry. We study the heterotic string on a factorizable six-dimensional orbifold which includes $T^2/Z_3$. The coordinates are denoted by $X^1$ and $X^2$. We start with the discussion of the two-dimensional torus, $T^2$, without orbifolding. There is always a $U(1)^2$ symmetry corresponding to the two currents, $H_1 = i\partial X^1$ and $H_2 = i \partial X^2$. At a certain point in the moduli space of $T^2$, there appear additional six massless gauge bosons. Then, the gauge symmetry is enhanced from $U(1)^2$ to $SU(3)$. The corresponding Kac-Moody currents are
\begin{eqnarray}
E_{\pm 1,0}, \qquad E_{0, \pm 1}, \qquad E_{\pm 1, \pm 1},
\end{eqnarray}
with 
\begin{eqnarray}
E_{n_1,n_2} = e^{i \sum_{i=1,2} (n_1 \alpha_1^{i} +n_2  \alpha_2^{i})X^i},
\end{eqnarray}
where $\alpha_1$ and $\alpha_2$ denote simple roots of $SU(3)$, i.e. $\alpha_1 = (\sqrt 2, 0)$ and $\alpha_2 = (-\sqrt 2/2, \sqrt 6/2) $. These currents, $H_i$ and $E_{n_1,n_2}$, satisfy the $su(3)$ Kac-Moody algebra.

Now, let us study the $Z_3$ orbifolding, 
\begin{eqnarray}
Z \rightarrow \omega^{-1} Z,
\end{eqnarray}
where $Z = X^1 + iX^2 $ and $\omega = e^{2\pi i /3}$. The currents $H_i$ and their linear combinations are not $Z_3$-invariant and the corresponding gauge symmetries are broken. On the other hand, two independent linear combinations of $E_{n_1,n_2}$ are $Z_3$-invariant and correspond to a $U(1)^2$ symmetry that remains on the $T^2/Z_3$ orbifold. Thus, the $SU(3)$ gauge group is broken to $U(1)^2$ by the orbifolding. It is convenient to use the following basis, 
\begin{eqnarray}
{H'}_1 &=&  \frac{i}{\sqrt 2} \left( E^1_1 - E^2_1 \right), \\
{H'}_2 &=&  - \frac{1}{\sqrt 2} \left(E^1_1 + E^2_1 \right),  \\
E'_{1,0} & = & \frac{1}{\sqrt 3} \left( i H_{\omega^{-1}} + E^1_{\omega^{-1}} + E^2_{\omega^{-1}} \right), \\
E'_{0,1} & = & \frac{1}{\sqrt 3} \left( i H_{\omega^{-1}} + \omega E^1_{\omega^{-1}} + \omega^{-1} E^2_{\omega^{-1}} \right), \\
E'_{-1,-1} & = & \frac{1}{\sqrt 3} \left( i H_{\omega^{-1}} + \omega^{-1} E^1_{\omega^{-1}} + \omega E^2_{\omega^{-1}} \right), \\
E'_{-1,0} & = &\frac{1}{\sqrt 3} \left( - i H_{\omega} + E^1_{\omega} + E^2_{\omega} \right), \\
E'_{0,-1} & = &\frac{1}{\sqrt 3} \left( - i H_{\omega} + \omega E^1_{\omega} + \omega^{-1} E^2_{\omega} \right), \\
E'_{1,1} & = &\frac{1}{\sqrt 3} \left( - i H_{\omega} + \omega^{-1} E^1_{\omega} + \omega E^2_{\omega} \right), 
\end{eqnarray}
where 
\begin{eqnarray}
H_{\omega^{-1}} &=& \frac{1}{\sqrt 2}\left( H_1 + iH_2 \right), \\
H_{\omega} &=& \frac{1}{\sqrt 2}\left( H_1 - iH_2 \right), \\
E^1_{\omega^{-k}} &=& \frac{1}{\sqrt 3}\left( E_{1,0} + \omega^k E_{0,1} + \omega^{-k} E_{-1,-1} \right), \\
E^2_{\omega^{-k}} &=& \frac{1}{\sqrt 3}\left( E_{-1,0} + \omega^k E_{0,-1} + \omega^{-k} E_{1,1} \right).
\end{eqnarray}
The $E'_{n_1,n_2}$ correspond to states with charges $(n_1 \alpha^1_1 +n_2 \alpha^1_2, n_1 \alpha^2_1 +n_2 \alpha^2_2)$ under the unbroken $U(1)^2$. They transform under the $Z_3$ twist action as follows:
\begin{align}
E'_{-1,0} & \rightarrow  \omega E'_{-1,0},\qquad  E'_{0,-1} \rightarrow \omega E'_{0,-1},\qquad E'_{1,1}  \rightarrow \omega E'_{1,1},  \nonumber \\
E'_{1,0} & \rightarrow  \omega^{-1} E'_{1,0},\qquad E'_{0,1}  \rightarrow  \omega^{-1}  E'_{0,1},\qquad E'_{-1,-1}  \rightarrow \omega^{-1} E'_{-1,-1}.
\end{align}
Thus, the first three $E'_{n_1,n_2}$ correspond to untwisted matter fields with charges $-\alpha_1, -\alpha_2$ and $\alpha_1+\alpha_2$ under the unbroken $U(1)^2$. We denote them as $U_1, U_2$ and $U_3$, respectively. The other three are their CPT conjugates. 
In addition, there are other untwisted matter fields $U$ which have vanishing $U(1)^2$ charges, but are charged under an unbroken subgroup of $E_8 \times E_8$.

Now, since the primed currents fulfill the same OPEs as their unprimed counterparts, it is justified to introduce bosons ${X'}^i$, so that
\begin{eqnarray}
{H'}^i &= & i \partial {X'}^i \\ 
E'_{n_1,n_2} & = & e^{i \sum_{i=1,2} (n_1 \alpha_1^{i} +n_2  \alpha_2^{i}){X'}^i}.\nonumber
\end{eqnarray}
The $Z_3$ twist action on $X^i$ can then be realized as a shift action on ${X'}^i$ as 
\begin{align}
{X'}^i \rightarrow {X'}^i + 2 \pi \frac{ \alpha^i_1}{3}.
\end{align}

In the twist representation there are three fixed points on the $T^2/Z_3$ orbifold, to each of which corresponds a twisted state. The two-dimensional bosonic string with the $Z_3$ boundary condition has a contribution of $h=1/9$ to the conformal dimension. As in the previous one-dimensional case, the twisted states can be described in the shift representation as follows. The left-moving momentum modes $p_L$ of the torus-compactified $SU(3)$ model lie on the momentum lattice
\begin{align}
\Gamma_{SU(3)}  \; \cup \; (\Gamma_{SU(3)} + w_1)  \; \cup \; (\Gamma_{SU(3)} - w_1),
\end{align}
where $\Gamma_{SU(3)}$ denotes the $SU(3)$ root lattice which is spanned by the simple roots of $SU(3)$, $\Gamma_{SU(3)} \equiv n_1 \alpha_1 + n_2 \alpha_2$, and $w_1 = ( \sqrt{2}/2, \sqrt{6}/6 )$ is the fundamental weight corresponding to $\alpha_1$. Then, the momenta $p_L$ in the $k$-shifted sector lie on the momentum lattice shifted by the $Z_3$ shift vector $s = \alpha_1/3$, 
\begin{eqnarray}
(\Gamma_{SU(3)} + k \frac{\alpha_1}{3}) \; \cup \; (\Gamma_{SU(3)} + w_1+ k \frac{\alpha_1}{3}) \; \cup \; (\Gamma_{SU(3)} - w_1+ k \frac{\alpha_1}{3}).
\end{eqnarray}
For $k=1$, there are three ground states with $p_L \in \lbrace \alpha_1/3, \alpha_2/3, -(\alpha_1+\alpha_2)/3 \rbrace$. They correspond to (would-be-massless) matter fields which we denote by $M_1$, $M_2$ and $M_3$, respectively. These matter fields are shown in Table \ref{Tab:FieldZ3}.
The states for $k=-1$ correspond to CPT-conjugates. As expected, the shifted ground states have conformal dimension $h = p_L^2/2 = 1/9$, which coincides with the twisted ground states. Indeed, the shifted states are related to the twisted states by a change of basis \cite{Dijkgraaf:1987vp}. The shifted states have definite $U(1)^2$ charges. 

\begin{table}[t]
\begin{center}
\begin{tabular}{|c|c|c|c|c|c|}
\hline
Sector & Field & $U(1)^2$ charge & $Z_3^2 $ charge  \\
\hline
\hline
U & $U$ & $(0,0)$  & $(0,0)$  \\  
\hline
U & $U_1$ & $-\alpha_1$  & $(0,0)$  \\  
\hline
U & $U_2$ & $-\alpha_2$  & $(0,0)$  \\  
\hline
U & $U_3$ & $\alpha_1 + \alpha_2$  & $(0,0)$  \\  
\hline
T & $M_1$ & $ \frac{\alpha_1}{3}$  & $(\frac{1}{3}, \frac{1}{3})$ \\  
\hline
T & $M_2$ & $ \frac{\alpha_2}{3}$  & $(- \frac{1}{3}, 0)$    \\  
\hline
T & $M_3$ & $ -\frac{\alpha_1 + \alpha_2}{3}$  & $(0, - \frac{1}{3})$  \\  
\hline
\end{tabular}
\caption[smallcaption]{Field contents of $U(1)^2 \rtimes S_3$ model from $Z_3$ orbifold. $U(1)^2$ charges are shown. Charges under the $Z_3^2$ unbroken subgroup of the $U(1)^2$ group are also shown.}
\label{Tab:FieldZ3}
\end{center}
\end{table}

From Table \ref{Tab:FieldZ3}, it turns out that the (would-be-massless) matter contents possess a $S_3$ permutation symmetry. Let $S_3$ be generated by $a$ and $b$, with $a^3 = b^2 = (ab)^2 = 1$. Then, for a point $(q_{1}, q_{2})$ on the two-dimensional $U(1)^2$ charge plane, $a$ and $b$ shall act as
\begin{align}
a:\begin{pmatrix}
 q_{1} \\
 q_{2} \\
\end{pmatrix} 
&\to
\begin{pmatrix}
- \frac{1}{2} & \frac{\sqrt{3}}{2} \\
- \frac{\sqrt{3}}{2} & - \frac{1}{2} \\
\end{pmatrix}
\begin{pmatrix}
 q_{1} \\
 q_{2} \\
\end{pmatrix}, \\
b:\begin{pmatrix}
 q_{1} \\
 q_{2} \\
\end{pmatrix} 
&\to
\begin{pmatrix}
1 & 0\\
0 & -1 \\
\end{pmatrix}
\begin{pmatrix}
 q_{1} \\
 q_{2} \\
\end{pmatrix}  
.
\end{align}
The action of $a$ is equivalent to the replacement $\alpha_1 \to \alpha_2 \to -(\alpha_1+\alpha_2) \to \alpha_1$. Then, the spectrum is left invariant if at the same time we transform the fields $F_i = (U_i, M_i)$ as $F_1 \to F_2 \to F_3 \to F_1$. The action of $a$ on the $F_i$ is described by the $3 \times 3$ matrix
\begin{align} \label{S31}
&\left( 
\begin{array}{ccc}
 0& 0 & 1 \\
 1 & 0 & 0 \\
 0& 1 & 0 \\
\end{array} 
\right).
\end{align}
The action of $b$ corresponds to $\alpha_1 \leftrightarrow  \alpha_1$ and $\alpha_2 \leftrightarrow - ( \alpha_1 + \alpha_2 ) $, so simultaneously transforming $F_1 \leftrightarrow F_1$ and $F_2 \leftrightarrow F_3 $ results in a symmetry of the spectrum. This transformation corresponds to the matrix 
\begin{align} \label{S32}
&\left( 
\begin{array}{ccc}
 1& 0 & 0 \\
 0 & 0 & 1 \\
 0& 1 & 0 \\
\end{array} 
\right).
\end{align}
The $S_3$ symmetry just shown does not commute with $U(1)^2$. Rather, $S_3$ and $U(1)^2$ combine to semi-direct product $U(1)^2 \rtimes S_3$.

Next we shall consider the situation where our orbifold moves away from the enhanced point by taking a certain VEV of the K\"ahler moduli field $T$, which corresponds to $H_\omega$. The K\"ahler modulus can be described by the $U_i$ fields as
\begin{align}
T = \frac{1}{\sqrt{3}} ( U_1 + U_2 + U_3 ).
\end{align}
The deformation is realized by the following VEV direction,
\begin{align} \label{VEV}
\langle U_1 \rangle = \langle U_2 \rangle = \langle U_3 \rangle. 
\end{align}
Note that this VEV relation preserves the $S_3$ discrete symmetry generated by \eqref{S31} and \eqref{S32}. However, the $U(1)^2$ gauge symmetry breaks down to a discrete $Z_3^2$ subgroup due to the presence of the $M_i$ fields. The two $Z_3$ charges $(z_1, z_2)$ are determined by $U(1)^2$ charges $(u_1, u_2)$ as $z_1 = q_1/\sqrt{2} - q_2/\sqrt{6}, z_2 = q_1/\sqrt{2} + q_2/\sqrt{6} $. The $Z_3^2$ charges are listed in Table \ref{Tab:FieldZ3}. The $Z_3$ actions are described by
\begin{align} \label{Z31}
&\left( 
\begin{array}{ccc}
 \omega & 0 &0 \\
 0 & \omega^{-1} &0 \\
 0& 0&1 \\
\end{array} 
\right), \\
 \label{Z32}
&\left( 
\begin{array}{ccc}
 \omega & 0 &0 \\
 0 & 1 &0 \\
 0& 0& \omega^{-1} \\
\end{array} 
\right).
\end{align}
The matrices \eqref{S31}, \eqref{S32}, \eqref{Z31} and \eqref{Z32} are nothing but the generators of $\Delta(54) \simeq ( Z_3 \times Z_3 ) \rtimes S_3 $ in the ${\bf 3}_{1(1)}$ representation \cite{Escobar:2008vc}. Thus, the fields $(M_1, M_2, M_3)$ transform as the ${\bf 3}_{1(1)}$ under $\Delta(54)$, and the  field $U$ is the $\Delta(54)$ trivial singlet ${\bf 1}$. This reproduces the known properties of  ordinary $Z_3$ orbifold models at a general point in moduli space \cite{Kobayashi:2006wq}. Summarizing, the origin of the $\Delta(54)$ discrete symmetry in orbifold models can be explained as follows:
\begin{align} 
SU(3) \  \xrightarrow[{\rm orbifolding}]{} U(1)^2 \rtimes S_3 \  \xrightarrow[\langle T \rangle ]{}  \Delta(54) \ . 
\end{align}
There are other VEV directions that one might consider. For $\langle U_1 \rangle \neq \langle U_2 \rangle = \langle U_3 \rangle = 0$ the $U(1)^2 \rtimes S_3$ symmetry is broken to $(U(1) \rtimes Z_2) \times Z_6$. In the case where $\langle U_1 \rangle \neq \langle U_2 \rangle = \langle U_3 \rangle \neq 0$ one obtains $Z_3 \times S_3$. Finally, when all VEVs are different, i.e. $\langle U_1 \rangle \neq \langle U_2 \rangle \neq \langle U_3 \rangle \neq \langle U_1 \rangle$ the symmetry is broken to $Z_3 \times Z_3$.

\section{Conclusion}

We showed that non-Abelian discrete symmetries in heterotic orbifold models originate from a non-Abelian continuous gauge symmetry. The non-Abelian continuous gauge symmetry arises from torus-compactified extra dimensions at a special enhanced point in moduli space. In the two-dimensional orbifold case, by acting with $Z_3$ on the torus-compactified $SU(3)$ model, the non-Abelian gauge group $SU(3)$ is broken to a $U(1)^2$ subgroup. We observed that the matter contents of the orbifold model possess a $S_3$ symmetry which is understood to act on the two-dimensional $U(1)^2$ charge plane. The resulting orbifold model then has a symmetry of semi-direct product structure, $U(1)^2 \rtimes S_3$. In the untwisted sector, the orbifold model contains a K\"ahler moduli field which is charged under the unbroken Abelian gauge group. By assigning a VEV to the charged K\"ahler moduli field, the orbifold moves away from the enhanced point and the $U(1)^2$ gauge symmetry breaks to a discrete $Z_3^2$ subgroup. Thus, effectively the non-Abelian discrete symmetry $\Delta(54) \simeq ( Z_3 \times Z_3 ) \rtimes S_3$ is realized. The other VEV directions of the untwisted scalar fields break the symmetry to $(U(1) \rtimes Z_2) \times Z_6$, $Z_3 \times S_3$ or $Z_3 \times Z_3$. In the one-dimensional $Z_2$ orbifold case, we showed that the non-Abelian gauge symmetry $SU(2)$ is the origin of the discrete symmetry $D_4 \simeq Z_4 \rtimes Z_2$. The other VEV directions of the untwisted scalar fields break the symmetry to $Z_4$.

The resulting non-Abelian discrete flavor symmetries are exactly those that have been  obtained from heterotic string theory on symmetric orbifolds at a general point in moduli space \cite{Kobayashi:2006wq}. In \cite{Kobayashi:2006wq}, the geometrical symmetries of orbifolds were used to derive these discrete flavor symmetries. However, in this paper, we have not used these geometrical symmetries on the surface, although obviously the gauge symmetries and geometrical symmetries are tightly related with each other. At any rate, our results also indicate a procedure to derive non-Abelian discrete symmetries for models where there is no clear geometrical picture to begin with, such as in asymmetric orbifold models \cite{Narain:1986qm, Ibanez:1987pj, AsymmetricGUT, AsymModel} or Gepner models \cite{Gepner:1987qi}.

We give a comment on anomalies. Anomalies of non-Abelian discrete symmetries are an important issue to consider (see e.g. \cite{Araki:2008ek}). We start with a non-Abelian (continuous) gauge symmetry and break it by orbifolding and by moduli VEVs to a non-Abelian discrete symmetry. The original non-Abelian (continuous) gauge symmetry is anomaly-free and if it were broken by the Higgs mechanism, the remaining symmetry would also be anomaly-free. That is because only pairs vector-like under the unbroken symmetry gain mass terms. But this does not hold true for orbifold breaking, as it is possible to project out chiral matter fields. Thus, in our approach the anomalies of the resulting non-Abelian discrete symmetries are a priori nontrivial. However, in our mechanism we obtain semi-direct product structures such as $U(1)^2 \rtimes S_3$. Since the corresponding $U(1)^2$ is broken by the Higgs mechanism, the remnant $Z_3^2$ symmetry is expected to be anomaly-free if the original $U(1)^2$ is anomaly-free (the semi-direct product structure automatically ensures cancellation of $U(1)$-gravity-gravity anomalies, but other anomalies have to be checked). Thus, the only discrete anomalies that remain to be considered are those involving $S_3$.

We also comment on applications of our mechanism to phenomenological model building. In our construction the non-Abelian gauge group is broken by the orbifold action. This situation could be realized in the framework of field-theoretical higher-dimensional gauge theory with orbifold boundary conditions. Furthermore, our mechanism indicates that $U(1)^m \rtimes S_n$ or $U(1)^m \rtimes Z_n$ gauge theory can be regarded as a UV completion of non-Abelian discrete~symmetries\footnote{See also \cite{Abe:2010iv}.}. Thus, it may be possible to embed other phenomenologically interesting non-Abelian discrete symmetries into such a gauge theory and investigate their phenomenological properties.


\subsection*{Acknowledgement}
F.B. was supported by the "Leadership Development Program for Space Exploration and Research" from the Japan Society for the Promotion of Science, and by the Grant-in-Aid for Scientific Research from the Ministry of Education, Science, Sports, and Culture (MEXT), Japan (No. 23104011). T.K. was supported in part by the Grant-in-Aid for Scientific Research No.~25400252 from the Ministry of Education, Culture, Sports, Science and Technology of Japan. S.K. was supported by the Taiwan's National Science Council under grant  NSC102-2811-M-033-008.


\end{document}